"Hadron thermodynamics, concavity and negative heat capacities"

J. Dunning-Davies,
Department of Physics,
University of Hull,
Hull HU6 7RX,
England.

J.Dunning-Davies@hull.ac.uk

**Abstract.**
Problems associated with the nonextensivity of a hadron gas were discussed some time ago. Recently the topic has come to light again and here earlier results are re-examined in the light of new knowledge and attention is drawn to problems with the more modern work.

Hadron thermodynamics deals with a gas of indistinguishable particles whose mass spectrum is taken to have the form $Km^a e^{bm}$. Some time ago [1], a long-standing inconsistency was revealed and corrected. That inconsistency amounted to the non-extensive nature of the expression for the entropy found in some treatments by way of the microcanonical ensemble contrasting sharply with the extensive nature found by way of the canonical ensemble. The former result was shown to be due to an error. After correction of that error, the two ensembles were found to lead to the same expressions for the thermodynamic quantities if $a \geq -7/2$. However, in the case when $a < -5/2$, the microcanonical approach was used to consider the situation when one particle was appreciably heavier than the rest. Due to the fact that the resulting entropy was not found to be superadditive, it was concluded that this resulting entropy expression was unphysical. However, since that time, it has emerged [2] that it is the property of concavity of the entropy which embodies the essence of the Second Law of Thermodynamics, not the property of superadditivity as used to be felt to be the case. Hence, that conclusion of the earlier article [1] must be in error since the discussion followed the lines of noting that the entropy expression was not extensive; then noting that it was, in fact, concave; and, by using an earlier result [3] that concavity plus superadditivity implies extensivity, making the deduction that the entropy must not be superadditive. Hence, it was shown quite clearly that for a hadron gas which had one particle appreciably heavier than the rest and for which $a < -5/2$, the entropy expression is concave and, therefore, since it is the property of concavity which is now known to embody the essence of the second law, this entropy expression must be physically acceptable.

It is this property of concavity which excludes the possibility of negative heat capacities in closed systems. It must be realised that that is all that it excludes. As has been pointed out previously [4], negative heat capacities in open systems are quite acceptable because, since the Euler relation is of the form

$$TdS = dU + pdV - \mu dN,$$

the expression for the heat capacity in an *open* system is of the form

$$C_V = T\left(\frac{\partial S}{\partial T}\right)_V = \left(\frac{\partial U}{\partial T}\right)_V - \mu\left(\frac{\partial N}{\partial T}\right)_V,$$

where $\mu$ and $N$ represent chemical potential and number of particles respectively. Since the second term on the right-hand side of this expression has a negative sign associated with it, the sign of the heat capacity itself must remain indeterminate. Hence, the heat capacity of an *open* system *could* be negative. This conclusion does not violate the Second Law. However, an *open* system cannot be isolated and, if such a system and its surroundings are in equilibrium and are considered together as a composite system, then that composite system will be a *closed* system possessing a positive heat capacity.

It is often the case when negative heat capacities are discussed that there is a tacit assumption that systems possessing such heat capacities are allowable and attention is confined to drawing conclusions based on this assumption. However, it is clear [4] that the existence of such heat capacities in closed systems could lead to violation of the Second Law of Thermodynamics; it is only in open systems that the existence of such heat capacities is permissible.

It might be noted also that the various ensembles used in statistical thermodynamics are merely mathematical tools. A particular ensemble is chosen for use in a particular case either because it corresponds closely to the experimental position being investigated or, possibly more usually, because it proves easier to use. However, the truly important item is the actual physical system under investigation. The results obtained by using a particular ensemble are only as good as their reflection of the behaviour of that system. Hence, if one ensemble leads to one set of conclusions but a second leads to a different set, it is the whole theoretical discussion that should come under review. Obviously, as pointed out by Hill [5], if systems are extensive, all ensembles lead to the same results. It is only when systems seem non-extensive that problems arise. If an entropy does turn out to be non-extensive, it was usual to restrict attention to results obtained via the ensemble most closely aligned with the physical detail of the system under discussion. The modern trend to use so-called 'Tsallis' statistics should be viewed with caution, as is evidenced by the article by Nauenberg [6] and the series of queries raised by Lavenda [7]. It must be remembered always that this 'Tsallis' entropy was not simply postulated in 1988, it had been known and used by information theorists for many years before that, with many of the results available in the book by Hardy, Littlewood and Pólya [8]. Hence, questions about use of this expression in physical problems obviously abound.

Finally, it might be noted that there are real problems associated with both string theory and also inflation as has been discussed elsewhere [9] and all this raises serious questions, which need to be answered, about some of the material in the recent article by Cobas, et al [10], where the existence of negative heat capacities for physical systems seems to be regarded as acceptable and the fact that ensembles are simply mathematical tools which enable the discussion of physical systems and no more, having little relevance when separate from those systems, seems overlooked.